\documentclass[prl,preprint,amsmath,amssymb,floatfix,eqsecnum]{revtex4}
\usepackage{graphicx}
\usepackage{bm}

\begin{document}

\title{Scratching the Bose surface}

\author{Subir Sachdev}
\email{subir.sachdev@yale.edu}
\homepage{http://pantheon.yale.edu/~subir} \affiliation{Department
of Physics, Yale University, P.O. Box 208120, New Haven CT
06520-8120}

\begin{abstract}
This is a `News and Views' article (Nature {\bf 418}, 739 (2002))
discussing recent proposals for ground states of many boson
systems which are neither superfluids nor Mott insulators.
\end{abstract}

\maketitle

{\bf There are two distinct types of particles in nature: fermions
and bosons. But it seems bosons may assume similar characteristics
to fermion systems in the low-temperature regime typical of
Bose-Einstein condensation.}

Particles known as fermions (such as electrons) obey the Pauli
exclusion principle, which decrees that only a single fermion can
occupy a particular state, such as an orbital in an atom. But the
second family of particles, bosons (such as helium atoms), have no
corresponding restrictions on their occupation of states. This
distinction in their individual properties also carries over to
the collective properties of large numbers of fermions or bosons
at low temperatures. The typical quantum state of many fermions is
a Fermi liquid, formed, for example, by the valence electrons in
all metals: the electrons can move from atom to atom in this state
and conduct electricity across macroscopic distances, albeit with
a finite resistance. In contrast, bosons typically form a
Bose-Einstein condensate, which is the basis of superfluidity in
helium and its ability to flow perpetually with negligible
dissipation.

Now, Paramekanti and colleagues \cite{1}, in an article to appear
in Physical Review B, have proposed an innovative state for boson
systems which shares many of its characteristics with the metallic
state normally associated with fermions. According to Paramekanti
{\em et al.}, strong interactions among the bosons entangle them
in a delicate quantum-mechanical superposition that mimics the
properties of fermions.

One of the defining properties of a Fermi liquid is its Fermi
surface. In a metal, we can label the states available to the
fermionic electron by its momentum vector {\bf k}. The Fermi
surface resides in {\bf k}-space, and in the lowest-energy state
of the metal, electrons occupy all states inside the Fermi
surface, while those outside the surface remain unoccupied (Fig.
1a). This surface is of particular physical importance because it
defines the low-energy excitations observed experimentally: a
fluctuation in the density of the electrons is created by moving
an electron from an occupied state inside the Fermi surface to an
unoccupied state outside the Fermi surface; probes that eject an
electron from the metal (such as an energetic photon in a
photoemission experiment) will act most efficiently on states just
inside the Fermi surface. The Pauli exclusion principle, by
enforcing single occupancy of fermionic states, is crucial in
defining the concept of a Fermi surface, as electrons can only be
removed (or added) from states inside (or outside) the Fermi
surface.

The striking proposal of Paramekanti {\em et al.} is that bosons
can also form a Bose liquid with a Bose surface, analogous to the
Fermi liquid and Fermi surface. Clearly, as there is no Pauli
exclusion principle for bosons, the Bose surface cannot be the
boundary between occupied and unoccupied states. Instead, the Bose
surface defines the locus of points in momentum space with
low-energy excitations (Fig. 1b). Nevertheless, the excited states
of the Bose liquid are defined by the Bose surface in a manner
that is very reminiscent of the Fermi surface in a Fermi liquid:
low-energy fluctuations in the density of the bosons involve
rearrangement of the states near the Bose surface, as does an
excitation that ejects a boson from the liquid.

Paramekanti {\em et al.} also discuss the conditions under which
this fascinating Bose liquid may form. Weakly interacting bosons
invariably collapse into the single state of a Bose-Einstein
condensate, but merely turning up the strength of the repulsive
interactions between the bosons is not enough to establish a Bose
liquid. In fact, this leads to a state known as a Mott insulator,
in which the bosons localize in a regular, crystalline
arrangement. The transition between the Bose-Einstein condensate
and the Mott insulator in a trapped gas of rubidium atoms was
observed recently in a beautiful experiment by Greiner {\em et
al.} \cite{2}.

To deter the bosons from forming a Bose-Einstein condensate or a
Mott insulator, a more complex interaction between the bosons is
necessary. In particular, a large contribution from 'boson ring
exchanges' appears to be crucial. This feature arises because the
quantum-mechanical description---or wavefunction---of the ground
state contains a superposition of states in which pairs of bosons
move in a correlated manner around a ring. In such a state, the
positions of the bosons around the ring are uncertain, but a
measurement that locates one boson at a particular site also
specifies the position of the second boson at another site on the
ring.

Stimulated by the proposal of Paramekanti {\em et al.}, Sandvik
{\em et al.} \cite{3} have already reported results from a
computer simulation of the simplest quantum description of boson
behaviour in two dimensions, including a large boson-ring-exchange
term. So far, they have not found a state with a Bose surface, but
have instead obtained a new form of Mott insulator in which the
bosons crystallize on half the horizontal bonds of the lattice;
this type of state had also been predicted \cite{4}.

Paramekanti {\em et al.} \cite{1} also discuss the prospects for
experimental discovery of their Bose liquid state. They suggest
that such a state might be formed by pairs of electrons (which act
like composite bosons and are known as Cooper pairs) in the
cuprate compounds that superconduct at high temperatures. Mason
and Kapitulnik \cite{5} have recently reported an unexpected
regime of metallic conduction in a disordered thin film of
Mo$_{43}$Ge$_{57}$ in a magnetic field---the film is a
superconductor in zero field, and a Bose liquid of Cooper pairs is
an intriguing possibility for the metallic phase. And there are
other competing theories\cite{6,7,8}, involving a more fundamental
role for disorder in the film. It is clear that the resolution of
the puzzle set by Paramekanti {\em et al.} brings the prospect of
much exciting new physics.

\begin{figure}
\centerline{\includegraphics[width=6in]{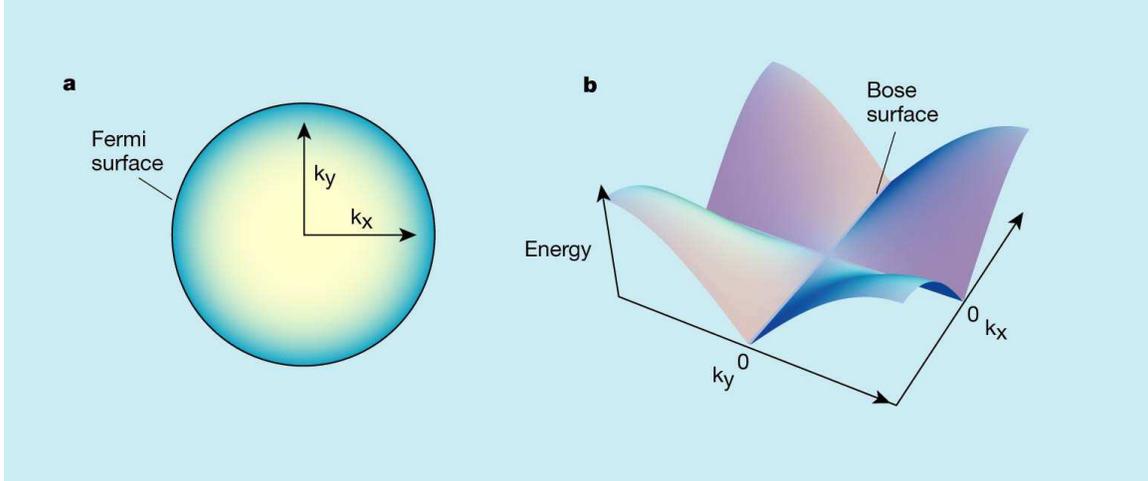}}
\caption{Fermi
and Bose surfaces in two dimensions. ({\em a\/}) Fermionic
electrons in a metal obey Pauli's exclusion principle--- there can
be only one fermion in each quantum state. The Fermi surface marks
the divide, defined by the particles' momentum components $k_x$
and $k_y$, between occupied and unoccupied momentum states. ({\em
b}) Although bosons do not obey the exclusion principle,
Paramekanti {\em et al.} \protect\cite{1} propose that they can
form a Bose surface in a Bose liquid, analogous to the Fermi
surface for fermions. In the spectrum of boson excitations in the
Bose liquid, the excitation energy (vertical axis) vanishes near
the Bose surface (which follows the lines $k_x = 0$ and $k_y =
0$): a Fermi liquid has a similar spectrum of excitations that
vanishes on its circular Fermi surface. } \label{fig1}
\end{figure}

\end{document}